\documentclass[twocolumn,twoside,prd,floatfix,letterpaper]{revtex4}

\usepackage{ifpdf}
\ifpdf 
  \usepackage[pdftex]{graphicx} 
\else 
  \usepackage{graphicx} 
\fi
\usepackage{amsmath}
\usepackage{amssymb}
\usepackage{fancyhdr}
\usepackage{color}
\usepackage{rotating}
\usepackage[colorlinks,hyperindex]{hyperref}

\begin{document} 

\title{How to Increase Global Wealth Inequality for Fun and Profit}
\author{Bruce Knuteson}
\noaffiliation

\begin{abstract}
We point out a simple equities trading strategy that allows a sufficiently large, market-neutral, quantitative hedge fund to achieve outsized returns while simultaneously contributing significantly to increasing global wealth inequality.  Overnight and intraday return distributions in major equity indices in the United States, Canada, France, Germany, and Japan suggest a few such firms have been implementing this strategy successfully for more than twenty-five years.
\end{abstract}

\maketitle

\section{Profit}
\label{sec:Profit}

The Strategy~\cite{knuteson2016information} is very simple:  construct a large, suitably leveraged, market-neutral equity portfolio and then systematically expand it in the morning and contract it in the afternoon, day after day.

The Strategy works because your trading will, on average, move prices in a direction that nets you mark-to-market gains.  Bid-ask spreads are wider and depths are thinner near market open than near market close~\cite{wsj2014intraday}, so aggressive trades early in the trading day move prices more than equally sized aggressive trades later in the day~\cite{cont2014price}.  An intraday round trip -- e.g., aggressively buying in the morning and selling in the afternoon -- thus nudges the market's midprice in the direction of your morning trading.  A reasonable level of daily round-trip trading~\footnote{We will sometimes shorten ``daily intraday round-trip trading'' to ``daily round-trip trading'' or ``intraday round-trip trading.''  In all cases we mean the same thing:  a round-trip trade performed in a single day (e.g., buying in the morning and selling in the afternoon), repeated day after day.  Any such repeated round-trip trade will probably need to be less than a few percent of the entire market's typical total daily volume.} combined with a sufficiently large portfolio will therefore produce expected mark-to-market gains exceeding the expected cost of your daily round-trip trading~\footnote{The cost (in local currency) of your daily round-trip trading depends on how much you choose to trade, but does not depend on the size of your existing equity portfolio.  Your expected mark-to-market gains (in local currency) will be proportional to the size of your existing equity portfolio.  With a sufficiently large portfolio, your expected mark-to-market gains (which are proportional to the size of your portfolio) will exceed the expected cost of your daily round-trip trading (which does not depend on the size of your portfolio).}. 

The availability of the Strategy of course depends on the practical threshold for ``sufficiently large.''  The only publicly available estimate of ``sufficiently large''~\cite{knuteson2016information} suggests the Strategy is available to institutions with roughly one billion dollars of available capital.

A few things should be kept in mind when running the Strategy.

The Strategy is illegal.  The spirit of the Strategy is unquestionably market manipulation, so regulators can shut you down and impose penalties if they wish.  Separately, taking money from outside investors without disclosing any ongoing or previous material use of the Strategy is fraud~\footnote{For example, you have committed fraud if (i) you claim or imply your previous favorable returns result from proprietary forecasts when (ii) you know or reasonably should know that your previous favorable returns accrued in significant part from the Strategy and (iii) you choose not to disclose your previous use of the Strategy to prospective investors (iv) who justifiably rely on your claim or implication that your previous favorable returns result from your proprietary forecasts and (v) who then subsequently lose money from their investment with you.  To be clear, you absolutely satisfy (i)--(iv) even if you previously and unknowingly used the Strategy and stopped the Strategy as soon as you realized you were using it, if you then subsequently took money from outside investors whose decision to invest was made in part on your track record of past returns and you did not disclose that your use of the Strategy may have contributed materially to your past returns.  We also note that the case of {\it{United States v. Shkreli}} establishes an interesting precedent for (v) being unnecessary.}.

You are unlikely to be caught.  Exposing Bernard Madoff's fraud~\footnote{Bernard Madoff operated the largest Ponzi scheme in history for well over a decade before his arrest in 2008.} merely required the United States Securities and Exchange Commission (SEC) to check whether Madoff had the money he claimed (he didn't) or was doing the trades he claimed (he wasn't).  The SEC did neither.  More recently and relevantly, we have been warning the SEC about the Strategy continually for years~\cite{knuteson2016information}.  This is the level of policing you can expect.

If other market participants are already following the Strategy, it is helpful if you can approximately align your portfolio with theirs.  This allows you to benefit from their round-trip trading, sharing the trading costs of marking up all of your books.  If explicitly telling each other your portfolios seems too dangerous, there are ways to make an intelligent guess.

Your actual trading can be quite complicated in detail while retaining the general pattern of systematically expanding your portfolio in the morning and contracting it in the afternoon.  Camouflaging your execution of the Strategy in this way costs more but significantly reduces your chances of getting caught.

The large bid-ask spread near market open and the day-to-day uncertainty in your trading prevents other market participants from fully arbing away the effect of your daily round-trip trading by, for example, selling at market open and buying at market close what you buy at market open and sell at market close.  The threat of such activity does limit your daily price nudge to less than or on the order of the spread near market open.  This limit enters the calculation of ``sufficiently large'' noted above.

You can execute the Strategy for a surprisingly long time.  Periodically raising additional capital allows you to further expand your book in a manner creating additional mark-to-market gains.  Slowly rotating your portfolio prevents you from ever needing to push the price of any particular stock to a patently absurd value.

You must ensure your execution of the Strategy benefits many people, reducing the incentive for people (including regulators) to look too closely.  We explain how to do this in Section~\ref{sec:IncreaseWealthInequality}.

\section{Increase wealth inequality}
\label{sec:IncreaseWealthInequality}

\begin{figure*}[p]
\includegraphics[width=7in]{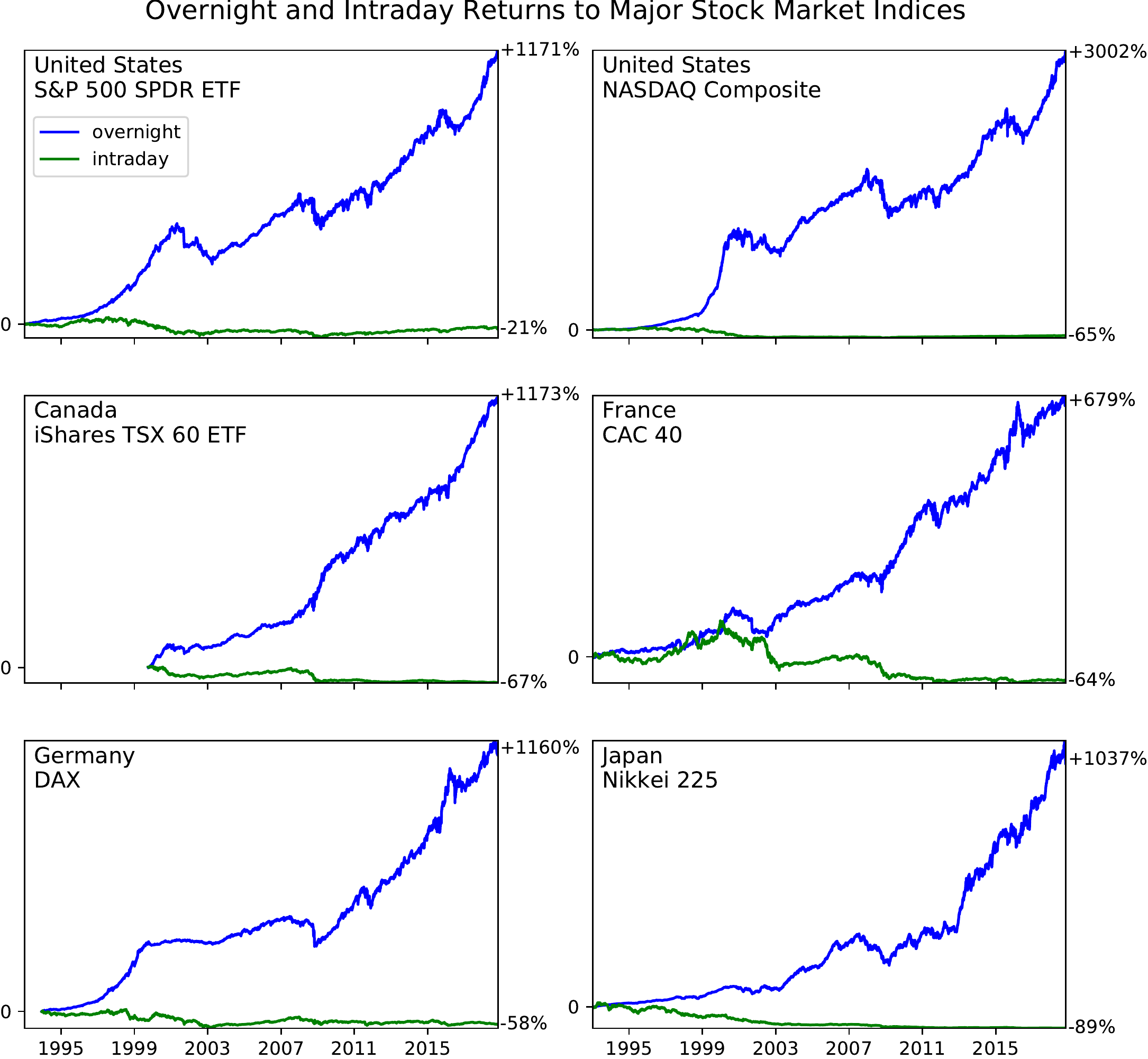}
\caption{\label{fig:StrategyFootprint}Cumulative overnight (blue curve) and intraday (green curve) returns to six major stock market indices over the past twenty-five years.  The overnight (blue) curve cumulates returns from market close to the next day's market open.  The intraday (green) curve cumulates returns from market open to market close.  The horizontal axis of each plot extends from  January 1, 1993 to October 31, 2018.  (Curves for the S\&P 500 SPDR ETF, the iShares TSX 60 ETF, and the DAX start on the first day for which data are available: January 29, 1993; October 4, 1999; and December 14, 1993, respectively.)  The (linear) vertical scale in each plot extends from a return of -100\% (bottom of plot) through 0 (explicitly marked, at left) to the largest cumulative overnight return achieved (top of plot).  On each plot, the cumulative overnight and intraday returns on October 31, 2018 are explicitly marked, at right.  The code used to make this figure is available at Ref.~\cite{thisArticleWebpage}.  Data are publicly available from Yahoo!~Finance.}
\end{figure*}

The key to the Strategy -- the intraday round-trip trading that systematically nudges prices in your favor -- is also your key to increasing global wealth inequality.

Even with a generally market-neutral book, your daily round-trip trading can create a drift in prices in the overall stock market~\footnote{A daily drift of merely 0.04\% -- well within the typical morning spread of most publicly traded stocks -- compounded over 25~years is the factor of 11 increase in the S\&P 500 index from 1993 to present.}.  If this drift in prices is downward, people will be angry, there will be congressional hearings, and you will be discovered quickly.  If the drift in prices is upward, people will be happy, there will be no congressional hearings, and you can run the Strategy for decades.  Make sure the drift in prices is upward.

The importance of this upward drift in overall prices is hard to overstate, as it grants you near immunity to serious investigation.  The key to any successful fraud is ensuring that anyone who knows enough to complain knows he is better off not complaining.  In this respect, the incentives created by the Strategy are impressive.

The money you lose on each of your daily round-trip trades (considered in isolation) feeds market makers.  You pay a commission for the use of a service, often provided by a large bank, that submits your order electronically to the exchange.  You pay a prime broker, again often a large bank, to finance your positions.  You pay a fee to the exchange itself for each transaction.  In the United States, you pay a transaction fee that funds the SEC~\footnote{More precisely: Congress sets the SEC's budget; the SEC periodically revises the value of the transaction fee so that the total expected fees collected are in line with the budget; and these transaction fees then fund the SEC.}.  The mark-to-market gains resulting from your use of the Strategy make money for your investors, and make you enough money to liberally compensate your employees.  To summarize, you directly pay every single entity in your immediate ecosystem that could conceivably have the data and expertise required to expose your use of the Strategy.

The incentives created by the Strategy outside your immediate ecosystem are no less impressive.  Every single individual with money in the stock market will be disinclined to believe the Strategy.  The Strategy is couched in the language of the price impact of orders, the quantitative details of which are not well understood by economists, and a topic on which the most active academic research is pursued primarily by a small group of former physicists~\footnote{The authors of Refs.~\cite{gatheral2010no,farmer2013efficiency,donier2015fully} are representative of the researchers most actively contributing to the public understanding of how orders determine long-term prices.}.  The question of how individual orders determine long-term prices remains publicly unresolved (although it has long been a solved problem in certain private circles), adding additional confusion to any public discussion (including the threshold of ``sufficiently large'' in Section~\ref{sec:Profit}).  Absent public clarity on this issue, the notion that a single market participant could profitably use the Strategy to affect overall stock market prices is {\it{prima facie}} absurd~\footnote{To the extent the existing literature addresses the predictable variation in price impact over the course of the trading day at all, it grossly underestimates both the magnitude of this variation and its persistence as impact decays.  The literature's lack of attention to this issue arises in part from an implicit decision to treat it as something to be patched onto a correct model of market impact, once achieved, rather than something so hard to patch that it must instead naturally fall out of the correct model.  The literature's significant underestimate of its magnitude arises in part from the analysis of collections of meta-orders that, both within individual meta-orders and across meta-orders, average over much of this variation.}, and few individual journalists will have the time, technical skill, and desire to probe further.  If any do, their efforts will be blocked further up their chain of command~\footnote{Few large news organizations these days are truly independent.  The financial conflicts of interest inherent in this particular story could hardly be greater.}.

Your round-trip trading may leave traceable footprints in trading records and return patterns.  These footprints need not be as obvious as Figure~\ref{fig:StrategyFootprint}, which shows cumulative overnight and intraday returns over the past 25~years in six major stock market indices:  the S\&P 500 index and the NASDAQ Composite index in the United States, Canada's TSX 60, France's CAC 40, Germany's DAX, and Japan's Nikkei 225~\footnote{The curves for the two ETFs include dividends with reinvestment, as one should.  The other four plots (NASDAQ Composite, CAC 40, DAX, and Nikkei 225) ignore dividends.  Properly including dividends with reinvestment in these four plots would increase the height of the blue curve in each plot and leave the green curve unchanged.}.  The return pattern in the S\&P 500 index was first pointed out over a decade ago~\cite{cooper2008return}.  Similar return patterns have been identified in major indices of other developed countries~\cite{lachance2015night}.  The only plausible explanation so far advanced for the highly suspicious return patterns in Figure~\ref{fig:StrategyFootprint} is someone using the Strategy~\cite{knuteson2016information}~\footnote{In United States equity markets, roughly two-thirds of a typical 24-hour-day's price variance realizes intraday and one-third realizes overnight.  Armed with this fact, the typical economist would predict the green curves to be roughly twice the height of the blue curves (since two-thirds is twice one-third).  Figure~\ref{fig:StrategyFootprint} looks nothing like this prediction, which fails to account for the predictable difference in bid-ask spread over the course of the trading day as an important limit to arbitrage and the outsized effect a few market participants executing the Strategy can have on overall price levels.  Shorting the overall market is of course possible in principle but cumbersome in practice, and anyone sophisticated enough to try is well aware that the market can remain irrational longer than she can remain solvent.}.

If you choose to execute the Strategy, Figure~\ref{fig:StrategyFootprint} suggests you will be joining other firms who have been employing it successfully for over twenty-five years.  The identity of these other firms is not for us to say, but the strikingly consistent nature of the plots in Figure~\ref{fig:StrategyFootprint} suggests firms whose trading is algorithmic~\cite{cooper2008return}~\footnote{Computers are consistent.  People are not.}.  It is therefore reasonable to guess that the firms whose trading is primarily responsible for Figure~\ref{fig:StrategyFootprint} are (i) quantitative trading firms that (ii) have been around since 1993 (iii) trading in volumes large enough to cause Figure~\ref{fig:StrategyFootprint} and (iv) with portfolios large enough to benefit from the Strategy.  The list of firms satisfying these criteria is not long.

We wish to emphasize again the extraordinary power and pervasiveness of the broad incentives created by proper use of the Strategy.  This is no conspiracy, where many people have malicious intent.  Nobody beyond your inner circle need have any malicious intent; the ability of any individual to call you out is naturally balanced by his financial incentives to remain silent; and any individual headstrong enough to complain is unlikely to find a receptive ear.  As with Madoff's fraud, the incentive structure here is masterful, even if it is more stumbled upon than designed.

Returning to the main point of this section, the manner in which you can significantly increase global wealth inequality should now be clear.  The Strategy, executed systematically and unchecked over a period of years, can systematically inflate an asset bubble.  In the United States, artificially doubling the price of publicly traded stocks creates over ten trillion dollars out of thin air, distributing it roughly in proportion to the wealth people already have.  Knock-on effects probably further amplify this direct effect~\footnote{Such knock-on effects seem to include, for example, an increased concentration of wealth into and within the financial and real estate sectors.}.  In countries with well-developed public markets, such as those named in Figure~\ref{fig:StrategyFootprint}, the ability to inflate a large asset bubble can thus overcome the most egalitarian system of social policies.  Even if your country suddenly embraces a number of policies designed to reduce wealth inequality~\footnote{For definiteness, suppose these policies include the full list from Ref.~\cite{stiglitz2018}: more progressive taxation, high-quality federally funded public education, affordable access to universities, strong anti-trust laws, more protective labor laws, executive salary caps, strict banking regulations, strong enforcement of anti-discrimination laws, higher estate taxes, guaranteed access to health care, strengthened retirement programs, and affordable urban housing.}, your ability to inflate a sufficiently large asset bubble can ensure the continued increase of wealth inequality within your country and make your country still wealthier than those lacking the large, well-developed public markets required for the Strategy.  With respect to wealth inequality, the ability to inflate a bubble in a large market trumps almost everything else~\footnote{Although we sympathize with much of Ref.~\cite{stiglitz2018}, we disagree that ``[t]here is no magic bullet to remedy a problem as deep-rooted as America's inequality.''   The magic bullet in this case is calling out the Strategy and shutting down any firms whose execution of the Strategy turns out, after considered investigation, to have contributed significantly to inflating the twenty-five-year-old asset bubble that is a primary cause of America's increased wealth inequality over the past quarter century.  Although obviously far from a complete solution to such an extraordinarily messy problem, this is as close to a magic bullet as you could possibly hope to find.}.

\section{Have fun}
\label{sec:Fun}

\setlength{\tabcolsep}{0.5em}
\begin{table*}[thb]
\begin{tabular}{llll}
{\bf Country } & {\bf Regulator} & {\bf Webpage} & {\bf Email } \\ \hline 
Canada & CSA & \url{https://securities-administrators.ca} &  csa-acvm-secretariat@acvm-csa.ca \\ 
France & AMF & \url{https://amf-france.org} & directiondelacommunication@amf-france.org \\ 
Germany & BaFin & \url{http://bafin.de} &  poststelle@bafin.de \\ 
Japan & FSA & \url{http://www.fsa.go.jp} &  equestion@fsa.go.jp \\ 
United States & SEC & \url{http://sec.gov} & chairmanoffice@sec.gov \\ 
\end{tabular}
\caption{Contact information for the regulators who should be able to definitively determine whose trading caused the highly suspicious return patterns (shown in Figure~\ref{fig:StrategyFootprint}) in the stock markets they oversee.}
\label{tbl:RegulatorEmailAddresses}
\end{table*}

If you or your institution has a billion dollars and want to make money while meaningfully increasing global wealth inequality, we hope you find the Strategy useful.  Have fun.

Alternatively, if you would like to know whose trading caused the highly suspicious return patterns shown in Figure~\ref{fig:StrategyFootprint}, you can send your local financial regulator a polite email asking if they would please take a look.  Appropriate email addresses are provided in Table~\ref{tbl:RegulatorEmailAddresses}.  If the cause of the highly suspicious return patterns in Figure~\ref{fig:StrategyFootprint} is what we think, emailing your financial regulator is the easiest and most effective single thing you can do to combat increasing global wealth inequality.

\bibliography{wi}

\end{document}